\documentclass[aps,preprintnumbers,amsmath,amssymb,11pt]{revtex4}

\usepackage{epsf}

\usepackage{graphicx}

\usepackage{bm}

\begin{document}


\def\Qs{Q_{\rm s}}

\def\half{{\textstyle{\frac12}}}
\def\p{{\bm p}}
\def\q{{\bm q}}
\def\x{{\bm x}}\def\v{{\bm v}}
\def\E{{\bm E}}
\def\B{{\bm B}}
\def\A{{\bm A}}
\def\a{{\bm a}}
\def\b{{\bm b}}
\def\c{{\bm c}}
\def\j{{\bm j}}
\def\n{{\bm n}}
\def\grad{{\bm\nabla}}
\def\da{d_{\rm A}}
\def\tr{\operatorname{tr}}
\def\Im{\operatorname{Im}}
\def\md{m_{\rm D}}
\def\mpl{m_{\rm pl}}
\def\pol{\varepsilon}
\def\bpol{{\bm\pol}}
\def\CP{CP^{N-1}}
\def\be{\begin{equation}}
\def\ee{\end{equation}}
\def\bea{\begin{eqnarray*}}
\def\eea{\end{eqnarray*}}


\title
    {
Topological insulators and the QCD vacuum: \\ the theta parameter as a Berry phase
    }

\author {H. B. Thacker\footnotemark \footnotetext{email: hbt8r@virginia.edu}
}
\affiliation
    {%
 Department of Physics,
    University of Virginia,
    P.O. Box 400714
    Charlottesville, VA 22901-4714\\
}

\date{\today}

\begin {abstract}%
{There is considerable evidence, based on large $N_c$ chiral dynamics, holographic QCD, and Monte Carlo studies,
that the QCD vacuum is permeated by discrete quasivacua separated by domain walls across which the local value of
the topological $\theta$ parameter jumps by $\pm2\pi$. This scenario is realized in a 2-dimensional U(1) gauge theory, the $CP^{N-1}$ sigma model,
where a pointlike charge is a domain wall, and $\theta$ describes the background electric flux and the polarization of
charged pairs in the vacuum. The transition between
discrete $\theta$ vacua occurs via the transport of integer units of charge between the two spatial boundaries of the domain.
We show that this screening process, and the role of $\theta$ as an order parameter describing electric polarization, are naturally formulated in terms of Bloch wave eigenstates of the Dirac Hamiltonian in
the background gauge field. This formulation is similar to the Berry phase description of electric polarization and quantized charge transport in topological insulators.
The Bloch waves are quasiperiodic superpositions of localized Dirac zero modes and the charge transport takes place coherently via topological charge-induced spectral flow.
The adiabatic spectral parameter becomes the Bloch wave momentum, which defines a Berry connection around the Brillouin zone of the zero mode band.
It describes the local polarization of vacuum pairs, analogous to 
its role in topological insulator theory. In 4D Yang-Mills theory, the $\theta$ domain walls are 2+1-dimensional Chern-Simons membranes, and the $\theta$ parameter
describes the local polarization of brane-antibrane pairs. The topological description of polarization in 2D U(1) gauge theory generalizes to membrane polarization in 4D QCD by exploiting
a relationship between the Berry connection and the gauge cohomology structure encoded in the descent equations of 4D Yang-Mills theory.
}%
\end {abstract}

\maketitle
\thispagestyle {empty}

\newpage
\section {Introduction}
In the modern theory of electric polarization in topological insulators \cite{Vanderbilt,Resta}, the concept of a Berry phase (sometimes called a
geometric phase) has played a central role. In this context, the Berry phase is the phase acquired by a charged-particle Bloch wave
state under adiabatic transport of the Bloch wave momentum around the Brillouin zone. This provides an understanding of
the topological origin of charge transport in such systems, in which the bulk material is an insulator with a mass gap, but current
is carried by special ``topologically protected'' edge states. In this paper, I will show that the Berry phase concept provides a natural
framework for discussing the role of the topological $\theta$ parameter in relativistic gauge theories such as QCD. In this formulation, the
$\theta$ parameter itself is a Berry phase, and the Bloch wave states that define it are constructed from the Dirac zero modes associated
with topological fluctuations of the gauge field. First we will discuss the interpretation of the $\theta$ parameter in two-dimensional U(1)
gauge theories (e.g. QED2 or $CP^{N-1}$ sigma models), where a $\theta$ term in the Lagrangian can be interpreted as a background electric field \cite{Coleman75}.
In this case, the $\theta$ dependence of the vacuum energy and nonzero topological susceptibility are determined by the polarizability of quark-antiquark pairs in the vacuum. 
For small values of the background electric field, these pairs do not come apart completely, but form localized dipoles which partially screen the background
field. But when the electric field exceeds a certain strength (corresponding to $\theta=\pm \pi$ or a half-unit of electric flux), it becomes energetically favorable for a pair to come apart,
with  charge ending up at $\pm$ spatial infinity (or on the capacitor plates that were erected to provide the background field). This screens off
exactly one unit of electric flux, leaving a net flux of a half-unit in the other direction. 
When the value of $\theta$ reaches $2\pi$, the bulk vacuum is back to having zero electric field. In the bulk, the 
$\theta=2\pi$ vacuum and the $\theta=0$ vacuum are indistinguishable. To distinguish between them, we must include the boundary charges on the capacitor plates.
This is quite analogous to the bulk-boundary constraint in topological insulators: the difference between the value of the bulk topological parameter $\theta$ before
and after a transition is determined by the amount of charge that has flowed from the bulk to the boundary. Current is only conserved if we include boundary charges.
In the context of relativistic gauge theories, the relation between bulk and boundary currents can be formulated covariantly as an ``anomaly inflow'' constraint \cite{TXK,Callan-Harvey}. 

The evidence for codimension-one Chern-Simons membranes as the dominant topological charge excitations in the QCD vacuum has been discussed previously \cite{Ahmad05}.
For 4-dimensional QCD, the operator which inserts such a membrane into the vacuum is the exponential of a 3D integral of the 3-index Chern-Simons tensor ${\cal K}_3$ over the surface of the membrane.
The topological charge density $Q(x)$ is obtained from the exterior derivative of this tensor, or equivalently, the 
divergence of the Chern-Simons current. 
For the case of 2D U(1) gauge theory, the topological charge density $Q=\frac{1}{2\pi}\partial^{\mu}K_{\mu}$ is proportional to the field strength,
\begin{equation}
Q= \frac{1}{2\pi}\epsilon_{\mu\nu}F^{\mu\nu} \equiv \frac{1}{2\pi}F
\end{equation}
so the Chern-Simons current $K_{\mu}$ 
is just the dual of the gauge potential $K_{\mu}=\epsilon_{\mu\nu}A^{\nu}$. For this case the integral over the membrane reduces to
an ordinary Wilson line integral. This allows us, in the 2D case, to interpret a Chern-Simons membrane as the space-time path of a point-like charged particle. 
Thus for example, in the 2D $CP^{n-1}$ sigma model,
the view of the vacuum as a condensate of tightly bound charged pairs, as suggested by both the large-N solution \cite{Witten79} and the lattice strong-coupling expansion \cite{Samuel},
is supported by the observed dominance of 1D coherent topological charge membranes in Monte Carlo configurations \cite{Ahmad05}, which are interpreted as the space-time paths of 
charged particle pairs in the condensate. The analogous structure of layered, codimension-one topological charge
membranes observed in 4-dimensional $SU(3)$ lattice gauge configurations \cite{Horvath03,Ilgenfritz} suggests a natural extension of these ideas to the actual theory of interest, 4-dimensional
QCD. 

In the 2D U(1) theories, the polarizability of charged pairs in the vacuum determines the 
topological susceptibility as well as the string tension, i.e. the strength of the long-range charge confining linear Coulomb potential. 
In this paper we will show how to formulate the connection between topology 
and polarization in 2D U(1) gauge theory in terms of a Berry connection
which relates the local value of $\theta$ to the polarization of vacuum pairs. In this respect it is quite similar to the treatment of polarization in 
topological insulators \cite{Vanderbilt,Resta}. 

The discussion of a Berry connection in relativistic gauge theory is facilitated by
considering the ``fermionic'' description of a gauge field configuration. In this approach, one characterizes any particular gauge configuration in terms of the spectrum and eigenfunctions 
of the associated Dirac operator. 
The fermionic description of gauge configurations is particularly useful for considering the structure of topological charge fluctuations in the gauge field.
The connection between topological gauge field fluctuations and the spectrum of the Dirac operator is embodied in the
the Atiyah-Singer index theorem, which relates the number of units of topological charge of the gauge field on a periodic Euclidean 2- or 4-torus 
to the net number (left minus right) of chiral zero eigenmodes of the associated Dirac operator. The connection between topological charge and Dirac zero modes is most directly formulated in Euclidean spacetime, 
but it is instructive to consider this connection in a Hamiltonian framework, in which it may be described in terms of spectral flow. In this description, rather than considering eigenmodes 
of the Euclidean Dirac operator, we choose a time direction and consider the eigenmodes of the Dirac Hamiltonian as a function of time. The number of units of topological charge on
the torus is related by the index theorem to the net number of left and right-handed chiral 
Hamiltonian eigenmodes which flow from positive to negative energy as we go once around the torus in the time direction. As I will 
discuss in this paper, the Hamiltonian spectral flow description of topological charge fluctuations can be recast in terms of a Berry phase. The evolution of the Hamiltonian eigenstates 
in time, which defines the spectral flow, is reinterpreted as an adiabatic evolution in a Bloch wave momentum variable that characterizes the Dirac Hamiltonian eigenstates. 
The Berry phase thus defined is a measure of the polarization of vacuum pairs, and it defines a local value of $\theta$. In the weak field perturbative limit there is no
dissociation of pairs, as depicted in Fig 1(a). This calculation 
reproduces the perturbative 
result for topological susceptibility from one-loop vacuum polarization \cite{Witten79}. For stronger fields, a transition between discrete vacua with $\theta$ differing by $\pm 2\pi$
is identified as a discrete shift of the Berry phase, representing the instantaneous charge transport caused by dissociation of each vacuum pair, with the members of each pair 
recombining with the opposite charged member of a neighboring pair. This is represented in Fig. 1(b). Just as in the theory of electric polarization \cite{Vanderbilt}, a nonzero Berry
phase represents the transport of units of charge across a cell of the spatial lattice which defines the Brillouin zone band structure of the wavefunctions. 
 
\begin{figure}
\vspace*{4.0cm}
\includegraphics{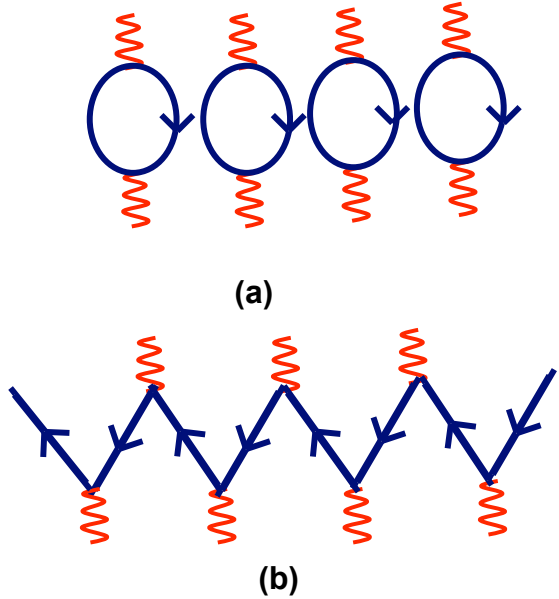}
\vspace{6.5cm}
\caption{(a) Polarization for small fields entails no net charge transport. (b) For larger fields pairs dissociate and annihilate with members of neighboring pairs, yielding a net
charge transfer between boundaries.}
\label{fig:polarization}
\end{figure}

\section{2-dimensional U(1) theory: topology and screening}

The relation between Hamiltonian spectral flow and the Berry phase description of topological charge is most easily seen in 2D U(1) theory for the case of a constant
field strength on a periodic 2-torus. Choosing $A_0=0$ gauge, the gauge interaction term in the Dirac Hamiltonian is $A_1 = Ft$, where $F\equiv \frac{1}{2}\epsilon^{\mu\nu}F_{\mu\nu}$ is the 
field strength and $t$ is Euclidean time. For a massless Dirac fermion in 2D Euclidean space, the Dirac eigenvalue equation separates into equations for left and right handed components:
\begin{eqnarray}
{\cal D}u_L(x,t)=\lambda u_L(x,t) \\
{\cal D}^*u_R(x,t)=\lambda u_R(x,t)
\end{eqnarray}
where
\begin{equation}
\label{eq:Dirac_op}
{\cal D} \equiv \frac{\partial}{\partial t} + i\frac{\partial}{\partial x} + Ft
\end{equation}
Let's first consider the equation for the left component.
We choose coordinates in the x-direction such that the period in that direction is $2\pi$, and we look for periodic solutions,
\begin{equation}
u_L(x+2\pi,t) = u_L(x,t)
\end{equation}
For field strength $F$ the topological charge density is $Q=F/2\pi$. For constant $F$ we have one unit of topological charge on the torus if we choose
the period in the time direction to be $T=1/F$. The Dirac wavefunctions are periodic in the spatial direction, but only quasiperiodic (periodic up to a gauge transformation) in the time direction,
\begin{equation}
u_L(x,t+T) = e^{ix}u_L(x,t)
\end{equation}
Because the integrated topological charge density is an integer, this gauge transformation is also periodic in $x\rightarrow x+ 2\pi$ so it is well-defined and continuous on the torus. 
For $F>0$, the index theorem implies the existence of a left-handed eigenmode
with zero eigenvalue, given explicitly by
\begin{equation}
\label{eq:zeromode}
u_L(x,t/T) = \sum_{n=-\infty}^{\infty} e^{-\frac{T}{2}(n+\frac{t}{T})^2} e^{inx}
\end{equation}
which satisfies ${\cal D}u_L=0$.
If the field strength $F$ is negative, there is instead a right-handed zero mode $u_R = u_L^*$. 

In a Hamiltonian framework, the index theorem manifests itself in the form of a spectral flow constraint. Consider the Dirac Hamiltonian as a parametric function of a rescaled time $k\equiv t/T$,
\begin{equation}
\label{eq:Hamiltonian}
H(t/T) = -i\frac{\partial}{\partial x} +\frac{t}{T}
\end{equation}
Acting on a periodic spatial interval $0<x<2\pi$, this Hamiltonian is periodic, up to a gauge transformation, over the Euclidean time interval $0<t<T$, i.e.
\begin{equation}
\label{eq:quasiperiodic}
H(k+1) = e^{-ix}H(k)e^{ix}
\end{equation}
Thus, for any integer $k$, the spectrum $E_n = n+k$ of $H(k)$ matches up with $E_n = n$, the spectrum of $H(0)$. The Dirac spectral flow for the gauge configuration is
given by $k$, the integer shift of the spectrum over the periodic time interval, or, more generally, the net number of left- minus right-handed modes that cross from negative
to positive energy. As first pointed out by Atiyah and Lustig \cite{Atiyah}, 
the spectral flow of the Dirac Hamiltonian is equal to the integrated topological charge over the 2D Euclidean space. 
Note that, if we quantize the Dirac Hamiltonian (as opposed to just employing the Dirac operator as a probe of the gauge field), 
spectral flow is associated with the production of fermion-antifermion pairs: a mode which crosses from negative to positive 
energy changes from filled to empty in the physical vacuum.  

In this example, the spectral flow parameter $k=A_1=Ft$ has a simple physical significance. It is the momentum imparted to the fermion from the background field over
the time interval $t$. This allows us to interpret the time evolution of the spectral flow as an adiabatic transport of the momentum parameter in the fermion
wave function, thus providing the essential elements of a Berry phase construction. 
Furthermore, the quasiperiodicity in Euclidean time (\ref{eq:quasiperiodic}) under an integer shift of $k$ reflects a Brillouin zone structure in momentum space. This is clarified
by rewriting the zero mode wavefunction via a Poisson transformation. The function (\ref{eq:zeromode}) is a Jacobi elliptic Theta function and a Poisson transformation on the sum corresponds
to a conjugate modulus transformation which effectively interchanges the the real and imaginary periods of the elliptic function. This gives
\begin{eqnarray}
\label{eq:Poisson1}
u^{(0)}_L(x,k) = \sum_{n=-\infty}^{\infty} e^{-\frac{T}{2}(n+k)^2} e^{inx} \\
\label{eq:Poisson2}
=\sum_{m=-\infty}^{\infty} e^{-\frac{1}{2T}(x-x_m)^2}e^{-ik(x-x_m)}
\end{eqnarray}
where we denote points of a spatial lattice by $x_m\equiv 2\pi m$. We also note that the zero mode wave function was constructed in the $A_0=0$ gauge. This function is
periodic in $x$, but only periodic up to a gauge transformation in the time direction. But the gauge transformation from the $A_0=0$ gauge to the $A_1=0$ gauge is just
\begin{equation}
g(x) = e^{i\frac{xt}{T}}= e^{ikx} 
\end{equation}
In the $A_1=0$ gauge, the wave function is periodic in time and quasiperiodic in space, and has the form of a Bloch wave,
\begin{equation}
\label{eq:Blochwave}
\Psi(x,k) = e^{ikx}u_L(x,t)
\end{equation}
This suggests interpreting the zero mode wave function $u_L$ on the torus $0<x<2\pi$ as the periodic part of a Bloch wave state on an infinite spatial lattice $x_m=2\pi m$. 
In the discussion thus far, we have regarded $F$ as a background electric field in one space and one time dimension. But the Euclidean 2D theory can just as well be interpreted
as a 2 space-dimensional system in a transverse magnetic field, as e.g. in the quantum Hall effect. Then a single periodic zero mode of the form (\ref{eq:zeromode}) can be interpreted as
a superposition of localized Landau orbitals. With a constant background field, the orbitals at different spatial locations are degenerate and regularly spaced, 
and can combine into delocalized Bloch wave states. This gives a natural lattice
structure, with lattice spacing determined by the radius of a Landau orbit and thus by the strength of the field $F$. Charge transport in such a system results from each
local orbit handing off a charge to the orbit to its left and picking up a charge from the orbit to its right as depicted in Fig. 1(b). The transport of charge across a single 
lattice cell is embodied in the phase of the periodic part of the Bloch wave, $u_L(x,k)$ in (\ref{eq:Blochwave}).

The interpretation of the $\theta$ parameter in 2D U(1) gauge theory as a local field describing the polarization of pairs in the vacuum can now be stated in terms of a Berry connection, 
following lines similar to the description of electric polarization in topological insulators \cite{Vanderbilt,Resta}. The Berry connection is constructed by differentiating
the periodic part of the Bloch wave function with respect to the momentum parameter:
\begin{equation}
\label{eq:Bconnection}
i\frac{\partial}{\partial k}u_L(x,k) = \sum_{m=-\infty}^{\infty} (x-x_m)e^{-\frac{1}{2T}(x-x_m)^2}e^{-ik(x-x_m)}
\end{equation}
The terms in the sum over $m$ represent individual Landau orbits localized around an origin at site $x_m$.
We see that differentiating by $k$ weights each term in the wavefunction by the dipole moment of the charge at $x$ {\it with respect to 
the center of its own Landau orbit.} (This can be taken as a charge-neutral dipole with the positive charge at $x$ and the negative charge at the origin $x=x_m$.)
The Berry connection ${\cal A}(k)$ at a particular lattice site is defined as the matrix element of
$i\frac{\partial}{\partial k}$ integrated over the cell,
\begin{equation}
\label{eq:berryconnection}
{\cal A}(k) = \int_{-\pi}^{\pi}dx\; u^*(x,k)i\frac{\partial}{\partial k}u(x,k)
\end{equation}
This plays the role of a gauge field in momentum space, where the phase change of the Dirac wave function under adiabatic translation in $k$ space
is given by the line integral of ${\cal A}(k)$.
The Berry phase which defines the local polarization field $\theta(x)$ is the gauge invariant closed loop integral around the lowest Brillouin zone $-\frac{1}{2}<k<\frac{1}{2}$,
\begin{equation}
\theta(x) = \int_{-\frac{1}{2}}^{\frac{1}{2}} {\cal A}(k) dk\equiv \oint {\cal A}(k)dk
\end{equation}
Here the range of integration $-\frac{1}{2}<k<\frac{1}{2}$ is over the Brillouin zone (BZ) associated with the lowest band of Bloch wave states, which are 
constructed from superpositions of left- and right-handed zero modes on a lattice cell. A gauge transformation of the Berry connection ${\cal A}(k)$ which
is topologically trivial over the BZ will leave the value of $\theta(x)$ invariant. On the other hand, a gauge transformation with nonzero winding number $n$ around
the BZ will change the value by $2\pi n$. These topologically nontrivial gauge transformations around the BZ represent the transfer of one or more units of charge 
between the two boundaries of the domain. They transform between vacua with values of $\theta$ which differ by integer multiples of $2\pi$. These vacua are
indistinguishable in the bulk (which is why $\theta$ is a truly periodic variable on a torus with no boundaries). For example, the $\theta=2\pi$ vacuum would have
one unit of background electric flux, but this is completely screened by the unit of charge that flowed to the boundary, so in the bulk there is zero electric field. 
Thus the value of $\theta$ keeps track of the current that has flowed to the boundaries. This is also the origin of the ``bulk-boundary'' constraint which relates
the change of the $\theta$ parameter across a membrane to the surface charge density on the membrane. This has been discussed previously in the context of anomaly inflow \cite{TXK}.

It is instructive to consider in more detail the adiabatic evolution of the Bloch wave states as we go around the BZ. We can write the time-dependent Dirac equation
for the periodic zero mode $u(x,k)$ as a formula for the adiabatic evolution of the Bloch wave in $k$ space,
\begin{equation}
\label{eq:adiabatic}
H(k)u(x,k) = \frac{1}{T}\frac{\partial}{\partial k}u(x,k)
\end{equation}
For large T, this equation traces out the adiabatic evolution of the eigenstates of $H(k)$ as a function of $k$. At any particular value of $k$, taking the $T\rightarrow \infty$
limit of $u(x,k)$ yields an eigenstate of $H(k)$. This follows from (\ref{eq:adiabatic}) but can also be seen from the explicit form of the zero mode $u_L(x,k)$ in (\ref{eq:Poisson1}). Using the first sum
in Eq. (\ref{eq:Poisson1}), the $T\rightarrow\infty$ limit is dominated by the nth term in the sum where n is the integer closest to $-k$ and the dominant eigenvector
is $u_n(x)\equiv e^{inx}$. This delineates the band structure of the spectrum. Each periodic eigenstate $u_n(x)$ on a single cell $0<x<2\pi$ with eigenvalue $E_n=n$ corresponds to a band of
Bloch wave states $k-\frac{1}{2}<n<k+\frac{1}{2}$ on an infinite lattice with eigenvalues $E_n(k)=n+k$. In particular, the lowest $n=0$ band of Bloch wave states $-\frac{1}{2}<k<\frac{1}{2}$,
is the zero mode band, with eigenvalues of $H(k)$ given by $E_0(k)=k$. In fact, in the physical vacuum, the $k<0$ states with negative energy are filled,
so the physical energy spectrum of the zero mode band is $E_0(k)=|k|, -\frac{1}{2}<k<\frac{1}{2}$. This has the important implication that this band can be regarded as a 
compact Brillouin zone by identifying the points $k=\frac{1}{2}$ and $k=-\frac{1}{2}$ where left- and right handed states are degenerate and can mix. 
Correspondingly, the adiabatic evolution across $k=0$ stays on the positive energy
branch and thus changes from Bloch waves built from $u_L(x,k)$ for $k>0$ to Bloch waves built from $u_R(x,k)=u_L^*(x,k)=u_L(x,-k)$ for $k<0$. The connectivity of the
positive energy branch across $k=0$ can be made more obvous by adding a small fermion mass term, which introduces a mass gap between the positive and negative
energy states and smooths out the cusp in $E_0(k)=|k|$.

We note that the evolution in k-space described by (\ref{eq:adiabatic}) becomes adiabatic in the limit where the time period $T\rightarrow\infty$. Since $T=1/F$, this corresponds
to the weak field limit $F\rightarrow 0$. In this limit, the spectrum obtained from all of the Bloch wave bands reproduces the full spectrum of 
a free massless Dirac fermion
\begin{equation}
E(k) = |k|,\;\; -\infty<k<\infty
\end{equation}
but separated into bands,
\begin{equation}
E_n(k) = |n+k|,\;\; -\frac{1}{2}<k<\frac{1}{2},\;\; n=0,\pm1,\pm2,\ldots
\end{equation}
From the Poisson transformed expression for the zero mode (the second sum in (\ref{eq:Poisson2})), we see that the gaussian wavefunctions centered around each lattice cell become completely
delocalized in the weak field limit $T\rightarrow\infty$. This is expected, since the radius of the Landau orbits is going to infinity in this limit. It is also instructive
to consider the opposite limit of strong field $T\rightarrow 0$. From the second sum in (\ref{eq:Poisson2}) we see that the cell wave functions in the Bloch wave become localized
around each lattice site $x_n$. In this limit, we expect that all of the Bloch wave states in a given band (i.e. all the states constructed from a particular periodic cell eigenstate
with different momentum $k$)
will be degenerate and separated by a finite gap from neighboring bands. This leads us to conclude that the qualitative effect of a finite background field $F$ is to introduce a mass gap 
which separates the $n=0$ band from the $n=\pm1$ bands at $k=\pm\frac{1}{2}$. It is the Berry phase around the BZ for the $n=0$ band that provides a local definition of the topological parameter
$\theta$. 

\begin{figure}
\vspace*{4.0cm}
\includegraphics{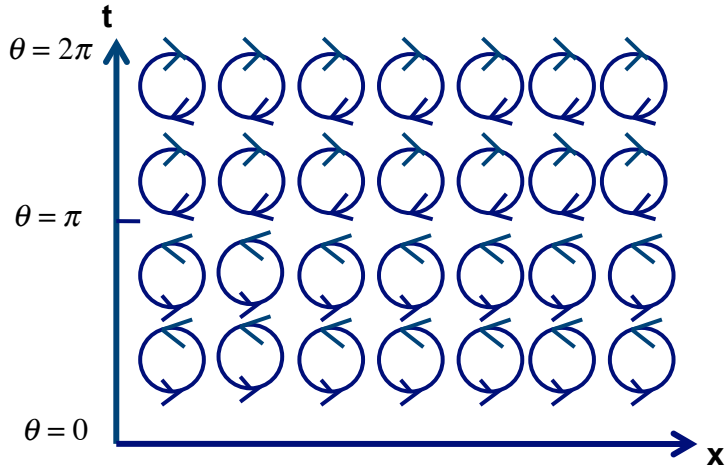}
\vspace{6.5cm}
\caption{The transition at the edge of the Brillouin zone from $k=+\frac{1}{2}$ to $k=-\frac{1}{2}$, which takes place at $\theta=\pi$, is accompanied by a chirality flip of
the Landau orbitals and the net flow of one unit of charge from boundary to boundary.}
\label{fig:chiralflip}
\end{figure}

The transition from $k>0$ to $k<0$ at the BZ boundary $k=\pm\frac{1}{2}$ has a simple physical interpretation. The strength of the field $F$ is proportional to $k$,
and the BZ boundary at $k=\frac{1}{2}$ corresponds to an applied field with a half-unit of flux. If we try to evolve adiabatically beyond the BZ boundary, the state suddenly 
shifts to $k=-\frac{1}{2}$. This indicates that one unit of charge has been transferred from the capacitor plate at $x=\infty$ to the one at $x=-\infty$. The net background 
field strength is now negative with a half-unit of flux. Since $F$ flips sign as we go from $k=+\frac{1}{2}$ to $k=-\frac{1}{2}$, the cell states flip from being left-handed (clockwise)
to right-handed (counterclockwise), as depicted in Fig. 2. In this Figure, the direction of the arrow represents the flow of electric current (dictated by the sign of $F$),
e.g. a left-handed mode, represented by a clockwise arrow, is either a clockwise-moving fermion or a counterclockwise-moving antifermion. Note that in Fig. 2, the transition from 
$k=\frac{1}{2}$ to $k=-\frac{1}{2}$ results in the charge flow between two adjacent rows of orbitals adding instead of cancelling. It thus 
entails a net flow of current from one end of the spatial volume to the other. As we continue the evolution in $k$, we go from $k=-\frac{1}{2}$
back to $k=0$, and the local electric field in the bulk returns to zero. In contrast, the Berry phase defined by the Bloch wave states we have constructed increases from 0 to
$2\pi$ as we go once around the BZ. For small values of applied field, $\theta$ is proportional to the local background electric field. But for larger applied field, 
mod $2\pi$ changes in the value of $\theta$ keep track of the number of charges that have flowed to the boundaries. 
In this way, the Berry phase definition of the local $\theta$ parameter for two-dimensional U(1) theory 
combines Coleman's interpretation of $\theta$ as a background electric field with the understanding
of quantized charge transport that is provided by topological insulator theory. 

\section{Yang-Mills cohomology and the Berry phase in 4D QCD}

In the preceeding section, we saw that the evolution of the Bloch wave under transport around the zero-mode BZ results in the net transfer of a unit of charge from one boundary 
to the other along the spatial axis.
This is the cumulative effect of the spectral flow on each unit cell. For a filled band, a complete cycle of the momentum $k$ around the compact BZ effectively transports the unit of charge 
at each site to the
left by one lattice cell $x\rightarrow x-2\pi$. On a given lattice cell, this charge transport is associated with the spectral flow. On each lattice cell, we need a gauge
transformation $g(x)=e^{-ik(x-x_m)}$ to recover periodicity in the time direction after evolution from $t=0$ ($k=0$) to $t=T$ ($k=1$). 
At any intermediate value of $k$, a gauge transformation by the Bloch wave phase, $g(x)=e^{-ikx}$,
transforms $A_1=k$ to $A_1=0$ but is not periodic in $x\rightarrow x+2\pi$. We note that the value of $k$ that appears in the Hamiltonian is just $A_1$, the spatial component of the vector potential.
For any fixed $k$, the additional gauge phase induced by going from $x_m$ to $x_{m+1}$ across a unit cell is $e^{i2\pi k}=e^{iA_1\Delta x}$. This is just the link phase, i.e. the
gauge group element which would be assigned to a link between two adjacent sites in a lattice formulation of the gauge field. The link phase is the gauge phase factor in the
fermion hopping matrix, representing the transport of a quark between neighboring sites. 

We can now understand the connection between a discrete $2\pi$ change of the Berry phase and the operation of 
inserting a Chern-Simons membrane into the vacuum via a Wilson line operator.
As discussed in the Introduction, a Chern-Simons membrane is created by a surface integral of the Chern-Simons tensor, which reduces in the 2D U(1) case to an ordinary Wilson line operator.
The Poisson transformed eigenfunction (\ref{eq:Poisson2}) may be understood as a superposition of quark-antiquark pair states originating from each lattice site $x_m$, forming an electric
dipole with moment $x-x_m$. (Here the quark is at $x$, while the antiquark is fixed at the lattice site $x_m$.) The phase factor in the $m$ th term of (\ref{eq:Poisson2}) is a gauge 
string, i.e. a Wilson line, from quark to antiquark,
\begin{equation}
e^{-ik(x-x_m)} = e^{i\int_x^{x_m}A_1dx'}
\end{equation}
The Berry phase is obtained from the phase associated with moving the quark from site $m$ to $m-1$,
\begin{equation}
\label{eq:theta}
\Theta(k) = \int_{x_{m-1}}^{x_m}A_1dx =2\pi k
\end{equation}
This phase defines a constant Berry connection 
\begin{equation}
{\cal A}(k) = \frac{\partial}{\partial k}\Theta(k) = 2\pi
\end{equation}
The Wilson line from $x_{m-1}$ to $x_m$ in (\ref{eq:theta}) can
be regarded as a closed loop around the spatial direction of the periodic torus representing a unit cell. In a time interval $t=kT$, this
loop sweeps out a fraction $k$ of the total surface area, and thus a fraction k of the total $F$ flux on the torus. 

To see how the Berry connection is generalized to 4D QCD, we replace the gauge potential in (\ref{eq:theta}) by the
Chern-Simons current and write the Berry phase as an integral of the Chern-Simons charge over the spatial cell,
\begin{equation}
\label{eq:theta2}
\Theta(k) = \int_{x_{m-1}}^{x_m}K_0\;dx =2\pi k
\end{equation}
Under a gauge transformationn $g=e^{i\omega}$, the variation of the CS current is
\begin{equation}
\label{eq:coho2}
\delta K_{\mu}=\epsilon_{\mu\nu}\partial^{\nu}\omega
\end{equation}
Thus the Berry phase also varies under a gauge transformation as,
\begin{equation}
\label{eq:gvar}
\delta\Theta(k) = \int_{x_{m-1}}^{x_m}\delta K_0 dx =\int_{x_{m-1}}^{x_m} \partial_1 \omega\; dx = \omega(x_m) - \omega(x_{m-1})
\end{equation}
For translation around a complete period of the BZ, $\Delta k=$ integer and the gauge transformation $\omega$
is periodic, so that $\theta$ can only change by a multiple of $2\pi$.
The change of electric polarization over the time period $T$ is given by the phase integral around the BZ
\begin{equation}
\Delta \theta(x) = 2\pi\Delta P=\oint {\cal A}(k)dk=\Theta(1)-\Theta(0)
\end{equation}
From (\ref{eq:gvar}) we see that the polarization is invariant for gauge transformations which are topologically trivial on the unit cell (i.e. $\omega(x_m)=\omega(x_{m-1})$).
A topologically nontrivial transformation (e.g. $g=e^{ix}$) will change the polarization by an integer. 

The extension of this definition of a Berry connection to 4-dimensional QCD is provided by the cohomological formulation of gauge topology first introduced
by Faddeev \cite{Faddeev84}. This approach is built on the descent equations of Yang-Mills theory.  
These equations define the interplay between gauge variations and exterior derivatives that relates the gauge variation of the Chern-Simons
3-form ${\cal K}_3$ to a WZW-type 2-form ${\cal K}_2$ which plays a role analogous to the gauge phase $\omega$ in the 2-dimensional case.
The equation we need follows from the gauge invariance of the topological charge, and the fact that it can be written as an exterior derivative of the 
CS 3-form. Using the standard notation of differential forms, the topological charge density $Q(x)$ is written in terms of the CS tensor ${\cal K}_3$ as
$Q = d{\cal K}_3$. We denote the gauge invariance of $Q$ by $\delta Q=0$, with $\delta$ representing an exterior derivative with respect to gauge parameter space.
This must commute with an exterior derivative in space-time, so
\begin{equation}
\delta Q = \delta(d{\cal K}_3) = d(\delta {\cal K}_3) = 0
\end{equation}
This means that the gauge variation of the CS tensor $\delta {\cal K}_3$ can be written {\it locally} as the exterior derivative of a 2-form 
\begin{equation}
\label{eq:coho4}
\delta {\cal K}_3 = d{\cal K}_2
\end{equation}
However, this equation is not valid globally. The WZW form ${\cal K}_2$ has the character of a phase, in that it is not single valued around topologically 
nontrivial closed paths.
In $SU(N_c)$ Yang-Mills theory, the topological charge is
\begin{equation}
Q(x) =\frac{1}{16\pi^2N_c} \varepsilon_{\mu\nu\sigma\tau}Tr F^{\mu\nu}F^{\sigma\tau}
\end{equation}
This is equal to the exterior derivative of the Chern-Simons 3-form, or equivalently, the divergence of the CS current $Q=\frac{1}{16\pi^2N_c}\partial^{\mu}K_{\mu}$, where
\begin{equation}
\label{eq:CScurrent}
K_{\mu} = \varepsilon_{\mu\alpha\beta\gamma}{\rm Tr}\left(A^{\alpha}\partial^{\beta}A^{\gamma} + \frac{2}{3}A^{\alpha}A^{\beta}A^{\gamma}\right)
\equiv \varepsilon_{\mu\alpha\beta\gamma}{\cal K}_3^{\alpha\beta\gamma}
\end{equation}
(Here and below, we drop the factor $\frac{1}{16\pi^2N_c}$ from our definition of the CS and WZW tensors.)
The WZW 2-form ${\cal K}_2$ is more complicated. It depends on both the gauge potential $A^{\mu}$ and the gauge phase $\omega = -i\log\;g$, where $g$ is the
$SU(N_c)$ color gauge transformation which defines the gauge variation in (\ref{eq:coho4}). Up to terms of 4th order in the gauge phase, it is
\begin{equation}
\label{eq:K_2A}
K_2^{\beta\gamma} = \frac{i}{3}{\rm Tr}\left[ \omega g^{-1}\partial^{\beta}g g^{-1}\partial^{\gamma}g \right] + 
{\rm Tr}\left[ \partial^{\beta}g\;g^{-1}A^{\gamma} \right] + {\cal O}(\omega^4)
\end{equation}

Cohomology amounts to a procedure for inverting the exterior derivative $d$ in (\ref{eq:coho4}) and determining solutions for ${\cal K}_2$ on a closed 3-surface from the gauge variation of the CS form ${\cal K}_3$.
Note that in the 2D U(1) version of Eq. (\ref{eq:coho4}), the role of the WZW tensor is played by the gauge phase $\omega$ itself, and the
descent equation is just the dual of the gauge transformation, i.e. Eq.(\ref{eq:coho2}).
The fact that $\omega$ need not be single valued around a closed loop, but only single valued mod $2\pi$, provides
the cohomology structure associated with topological charge quantization. Similarly, the WZW form $K_2$ involves the phase of the color gauge transformation $\omega = -i\log\;g$.
It's integral over a closed 3-surface is gauge invariant for small gauge transformations, but changes by integer multiples of $2\pi$ under topologically
nontrivial transformations.

In our discussion we have considered Dirac eigenmodes in a constant background field. The Bloch wave momentum parameter $k=t/T$ emerged as a rescaled
time variable in the Poisson transformed zero mode on a unit cell. Adiabatic transport in $k$ is obtained in the limit of large $T$. The Bloch wave structure
clarifies the coherent nature of charge transport and its relation to the spectral flow of Dirac zero modes. But our definition of the Berry connection is given locally
in terms of the charge transport across a single lattice cell.  If we allow nonperiodic gauge transformations on a unit cell, then any value of $A_1=k$ in the
Hamiltonian can be obtained from $k=0$ by a gauge transformation $g=e^{ikx}$. Physically, this describes the transport of a fractional amount of charge
to the boundary. This allows us to obtain the Berry phase directly from the descent equations by a construction that easily generalizes to 4-dimensional Yang-Mills
fields. For the 2-dimensional case we can write the Berry phase in terms of the gauge variation of the Chern-Simons charge $K_0$ integrated over the lattice cell,
\begin{equation}
\label{eq:thetak}
\Theta(k)-\Theta(0) = \int_{x_{m-1}}^{x_m}\delta K_0 \;dx = \omega(x_m)-\omega(x_{m-1}) = k(x_m-x_{m-1})
\end{equation}
This is the phase of a spatial Wilson line stretched between oppositely charged members of a polarized pair located at adjacent lattice sites $x_m$ and $x_{m-1}$.
A gauge transformation which varies $k$ can be regarded as a change of the physical distance between adjacent sites. The Berry phase is thus a local order
parameter which describes the polarization of vacuum pairs. 

As we have discussed in previous work \cite{Ahmad05,TXK}, a Wilson line in the 2D theory may be regarded
as a codimension-one Chern-Simons membrane, which generalizes in the 4D case to the integral of the Chern-Simons tensor over a 3-surface. The 4D analog of a charge
dipole is a polarized pair of CS membranes, and the Berry phase defines a local $\theta$ parameter describing the polarization of brane-antibrane pairs. As discussed previously
in the context of anomaly inflow \cite{TXK}, the gauge transformation on the brane surface that is implicit in the descent equations represents transverse fluctuations
of the branes. As in the 2D case, we consider the Dirac Hamiltonian on a 3-dimensional periodic cell in space. Choosing a gauge where $A_0=0$,
the time-dependent Schroedinger equation for the Dirac zero mode can be cast as a formula for adiabatic evolution of Hamiltonian eigenstates in the 3-momentum space of Bloch wave states
built from a lattice of unit cells. In the model of the QCD vacuum that seems to be favored by Monte Carlo results, 
the vacuum consists of a laminated stack of alternating sign membranes which are closely spaced in the direction transverse to the branes but flat and uniform over roughly the confinement scale in the
other three Euclidean directions. This strongly suggests that the Berry connection needed to define the $\theta$ parameter can be reduced to a one-dimensional Brillouin zone
describing Bloch wave momenta transverse to the branes. A fluctuation of $\theta$ thus describes the transverse polarization of the + and - topological charge membrane pairs.
From the point of view of quark eigenstates, the membrane polarization fluctuations described by $\theta$ represent Goldstone fluctuations of the chiral consensate.

We note that the spectral flow in k-space on a cell is implemented by a generally nonperiodic gauge transformation $e^{ikx}$ on the 
open line segment $x_m<x<x_{m+1}$ which zeros out the $A_1$ gauge link. In the 2D case,
the Bloch wave momentum is just given by the mismatch between gauge phases on opposite sides of the cell,
\begin{equation}
e^{ik(x_{m+1}-x_m)} = g(x_{m+1})g^{-1}(x_m)
\end{equation}
To explicitly construct $\Theta(k)$ for 4D QCD, we choose the x-axis to be transverse to the branes and consider, as in (\ref{eq:thetak}), the gauge variation of the CS charge 
$K_0$ integrated over a spatial unit cell that is a thin slab of space between two adjacent, oppositely charge branes at $x=x_m$ and $x=x_{m+1}$.
\begin{equation}
\label{eq:thetaberry}
\Theta(k)-\Theta(0) = \int_{x_{m}}^{x_{m+1}}\delta K_0\; d^3x = \Omega(x_{m+1})-\Omega(x_m)
\end{equation}
Here the gauge phases on the two boundaries are given by WZW 2-forms integrated over the 2-dimensional spatial surface of the branes. 
\begin{equation}
\Omega(x) = \int dy dz K_2^{yz}(A,\omega)
\end{equation}
The integrand depends on both the in-brane components of the color field and on the color gauge phase $\omega(x) = -i\log\;g(x)$. 
For a closely spaced brane-antibrane pair, a fluctuation of the relative color phase is equivalent to a fluctuation of the transverse gauge link between them. 
Expanding to lowest order in the brane separation,
\begin{equation}
\label{eq:Omega}
\Omega(x_{m+1})-\Omega(x_m) \approx k(x_{m+1}-x_m) 
\end{equation}
Here we have defined an effective Bloch wave momentum parameter
\begin{equation}
k = \frac{\partial \Omega}{\partial x}
\end{equation}
which determines the relative WZW phase on adjacent branes.
Just as in the 2-dimensional case, the phase difference across the unit cell represents the transport of quark charge from boundary to boundary. The transfer
of an integer number of quarks is described by a periodic but topologically nontrivial gauge transformation corresponding to integer valued $k$.
Equations (\ref{eq:thetaberry}) and (\ref{eq:Omega}) define a constant Berry connection, just as in the two-dimensional case,
\begin{equation}
\label{eq:bcon}
{\cal A}(k) = \frac{\partial}{\partial k}\Theta(k) = 2\pi
\end{equation}

Equation (\ref{eq:bcon}) defines the Berry phase for a topological transition which starts and ends with unpolarized brane-antibrane pairs. The positive branes have 
each moved over by one step and become coincident with the anti-brane from the next pair to the left. 
For more general connections, the integral of ${\cal A}(k)$ around the BZ defines $\theta(x)$, an order parameter describing the local polarization of the brane-antibrane pairs in the vacuum. 
A long wavelength polarization wave corresponds to a plane wave gauge transformation by a Bloch wave factor $e^{ikx}$ with small $k$.
For the 2D U(1) case, the Bloch factor is just the relative phase of the small gauge transformation $e^{ikx}$ on adjacent branes.
For 4D QCD the Bloch wave is defined by a slowly varying $SU(N_c)$ color gauge transformation $g(x)$. The $U(1)$ chiral phase $\theta(x)$ that 
describes the polarization of brane-antibrane pairs is determined from the color phase via the WZW 2-form evaluated on adjacent brane surfaces.. 

\section{Conclusion}

We have shown that the topological structure of the vacuum of QCD and the role of $\theta$ as a local order parameter can be formulated in terms of a 
Berry connection whose structure and physical significance is closely analogous to the theory of topological insulators. On a heuristic level, it is not
surprising that a Berry phase construction is the appropriate way to describe the short range topological structure of 
the QCD vacuum. A central feature of most applications of the Berry connection is the
existence of two distinct time scales in a problem, so that one can distinguish between ``fast'' and ``slow'' variables, with the slow variables being 
regarded as parameters in the Hamiltonian for the fast variables. A Berry connection can then be defined from the phase of Hamiltonian eigenstates 
under adiabatic change of the slow variables. For the vacuum of QCD, the fast variables are the short wavelength fluctuations of topological charge
in the direction transverse to the branes, where the sign of the charge alternates rapidly at the scale of the lattice cutoff. This short range structure is
explored by the Berry connection as we take the Bloch wave momentum around the BZ. In the continuum limit, what survives of the short range lattice structure
is the bulk polarization of the membranes, corresponding to small fluctuations of $\theta$, and the possibility of domain walls corresponding to boundaries
marking discrete changes of $\theta$ by $\pm 2\pi$. 

The relationship between topological charge membranes and the chiral condensate has been discussed in \cite{TXK,JLAB12}. The Berry phase construction provides some
additional insight into this connection. A well-known consequence of the axial U(1) anomaly is that a global change of $\theta$ is equivalent to a global chiral
rotation. The Berry connection defined by (\ref{eq:thetaberry}) extends this to a local relationship. 
Because of the axial U(1) anomaly, the flavor singlet axial-vector current which is conserved in the chiral limit
is not gauge invariant but rather varies in the same way as the Chern-Simons current under a color gauge transformation, $\delta j_5^{\mu}=\delta K^{\mu}$. 
In the gauge variation of the integral (\ref{eq:theta}), the CS charge $K_0$ can be replaced by the integral of the axial U(1) charge over the lattice cell, i.e. the generator
of a chiral rotation on the cell,
\begin{equation}
\label{eq:chiralberry}
\Theta(k)-\Theta(0) = \int_{x_{m}}^{x_{m+1}}\delta j^5_0\; d^3x = \Omega(x_{m+1})-\Omega(x_m)
\end{equation}
The Berry phase $\theta$ can thus be regarded as a local chiral rotation angle. 
In pure glue QCD without quarks, there are no physical massless Goldstone bosons. As discussed in (\cite{TXK}), a massless pole must appear in
the Chern-Simons current correlator, as required for nonzero topological susceptibility. However, this pole is cancelled by the massless pole in the $\partial_{\mu}\theta$ correlator in 
any gauge invariant amplitude. For 1-flavor QCD, the quark and antiquark in a pseudoscalar $\eta'$ meson are on an adjacent pair of oppositely charged branes, 
and the gauge link between them is modulated by the spacing between the branes. 
In this way, fluctuations of the membrane polarization $\theta$ generate Goldstone fluctuations of the chiral condensate. 
For the 1-flavor case, the $q\bar{q}$ Goldstone boson couples to the pure-glue CS current by quark-antiquark annihilation, which takes place via topologically
nontrivial brane fluctuations (since it requires the transport of a unit of quark charge between branes). This generates the $\eta'$ mass.
For two or more flavors of quark we can construct nonsinglet
Goldstone bosons with the quark and antiquark on adjacent branes carrying different flavors. With no $q\bar{q}$ annihilation, the long-wavelength fluctuations of
the brane polarization $\theta$ generate massless Goldstone pions. 

The picture of the chiral condensate which emerges from the Berry phase construction has
some similarity to the model of the chiral condensate in an instanton liquid \cite{Diakonov84}. In that model, the condensate is formed from the approximate
'tHooft zero modes of the Dirac operator on localized instantons and antiinstantons. The membrane model of the chiral condensate that we have discussed here shares
one aspect of the instanton liquid model, in that the condensate is formed from approximate topological Dirac zero modes. But in the instanton picture it is difficult to
understand long range propagation of massless Goldstone bosons in terms of localized `tHooft modes. By contrast, in the membrane model, the Dirac modes which form the condensate are surface
modes on the boundaries between discrete vacua. The delocalized nature of the membrane surface modes makes the long range propagation of Goldstone bosons much more plausible.

This work was supported by the Department of Energy under grant DE-SC00079984.


\begin {thebibliography}{}

\bibitem{Vanderbilt} 
R.~D.~King-Smith, and D.~Vanderbilt, Phys. Rev. B47, 1651 (1993).

\bibitem{Resta}
R.~Resta, Rev.~Mod.~Phys. 66,899 (1994).

\bibitem{Coleman75}
  S.~R.~Coleman,
  Annals Phys.\  {\bf 101}, 239 (1976).

\bibitem{TXK}
  H.~B.~Thacker, C.~Xiong and A.~Kamat,
  Phys.\ Rev.\ D {\bf 84}, 105011 (2011)
  [arXiv:1104.3063 [hep-th]].

\bibitem{Callan-Harvey}
C. Callan and J. Harvey, Nucl.~Phys. B250, 427 (1985).

\bibitem{Ahmad05}
S.~Ahmad, J.~T.~Lenaghan and H.~B.~Thacker, Phys.\ Rev.\ D72: 114511 (2005).

\bibitem{Witten79}
E.~Witten, Nucl. Phys. B149: 285 (1979).

\bibitem{Samuel}
S.~Samuel, Phys.~Rev.~D28: 2628 (1983).

\bibitem{Horvath03}
I.~Horvath et al., Phys. Rev. D68: 114505 (2003);.

\bibitem{Ilgenfritz}
E.~Ilgenfritz, et al., Phys. Rev. D76: 034506 (2007).

\bibitem{Atiyah}
M.~F.~Atiyah, Lecture Notes in Physics, vol. 208, Springer-Verlag (1984), 313.

\bibitem{Faddeev84}
L.~D.~Faddeev, Phys. Lett. 145B: 81 (1984).

\bibitem{JLAB12}
H.~B.~Thacker, PoS CD12: 060 (2013) [arXiv:1302.0535 [hep-th].

\bibitem{Diakonov84} 
D. Diakonov and V. Petrov, Phys. Lett. B147, 351 (1984).









































\end {thebibliography}
\end {document}